**Spin-ice dynamics of pyrochlore $Dy_2GaSbO_7$ with enhanced Pauling zero-point entropy**

S. Nandi,[a] Y.M. Jana,[a] J. Alam,[a] P. Bag,[b,c] S.S. Islam,[b] R. Nath[b]

[a] *Department of Physics, University of Kalyani, Kalyani 741235, Nadia, W.B., India*
[b] *School of Physics, Indian Institute of Science Education and Research, Thiruvananthapuram 695016, Kerala, India*
[c] *Department of Physics, National Dong Hwa University, Da Hsueh Road, Shoufeng, Hualien 97401, Taiwan*

**ABSTRACT**

We report the low-temperature magnetothermal properties and spin dynamics of a mixed pyrochlore, $Dy_2GaSbO_7$ through the coordination of measurements of dc magnetization, ac susceptibility, and heat capacity, and CF computation. In $Dy_2GaSbO_7$, the spins freeze at temperature $T_{ice} \approx 3.1$ K, corresponding to dynamical spin-freezing into a disordered state in compliance with the spin-ice rules, but with enhanced zero-point entropy, unlike cannonical dipolar spin-ice materials, $Dy_2Ti_2O_7$ and $Ho_2Ti_2O_7$, due to random disorder and varied chemical pressure of B-site (GaSb tetrahedra) ions.

# I. INTRODUCTION

Geometrically frustrated magnetic systems, in which the topology of the spin lattice leads to a frustration of the spin-spin interactions, have drawn continuing research interest because of the realization of a wide variety of novel low-temperature quantum states [1,2], e.g., spin-ice [3,4], spin-liquid-like, and spin-glass states, unconventional long-range-ordered and exotic low-energy excitations [5]. Pyrochlore oxides, having general formula unit (f.u.) $A_2B_2O_7$, where A is the trivalent rare-earth ($R^{3+}$) ion and B tetravalent transition metal ion, are the robust examples of frustrated system. These materials contain an infinite network of corner-sharing tetrahedra where the magnetic $R^{3+}$ ions (spins) reside at the corners of tetrahedra and hence cannot satisfy pair-wise AF interactions simultaneously. The exotic magnetic ground state is, therefore, achieved in pyrochlores by three competing interactions: nearest-neighbor (n.n.) AF exchange interactions, FM long-range dipolar interactions and directional anisotropies of crystal-field (CF) interaction, along with several other weak perturbations, e.g., further n.n. exchange interactions, thermal and/or quantum fluctuations ('order-by-disorder' mechanism) [6], external pressure [7], random disorder in the form of off-stoichiometry or intersite mixing [8], which have dramatic effects on the low-temperature properties of these systems.

Of particular interest, spin ices, which are directly mapped onto common 'water ice', have emerged as exemplary of highly frustrated systems in three dimensions [3,9,10]. In these systems, the spins are arranged in Pauling's "two-spins in, two-spins out" configurations onto the tetrahedron by the effective ferromagnetic interactions in the presence of strong CF induced Ising-type anisotropy. As a result, the magnetic systems are highly degenerate in the ground state in macroscopic scale and enter into a frozen disordered with a finite residual ('zero-point') entropy $S_0 = 1/2\ R\ \ln(3/2) = 1.68$ J $K^{-1}mol^{-1}$ below spin-freezing temperature. Consequently, the entropy recovered at high temperature should fall short of that expected for a two-level pseudo-spin-1/2 system [$S = k_B \ln(2^N) = R \ln 2$], resulting in a total entropy $S_P = 0.71R\ln 2 = 4.1$ J $K^{-1}mol^{-1}$ [1,10,11]. $Dy_2Ti_2O_7$ [11], $Dy_2Sn_2O_7$ [1,12] and also $Ho_2Ti_2O_7$ [13,14] are categorized as 'dipolar spin-ices' (DSI), in which FM n.n. dipolar interactions dominate over the weak AF exchange interactions [10,15] among the magnetic moments, oriented along the local <111> Ising directions of tetrahedral units by the trigonal CF. However, recent heat capacity measurement on $Dy_2Ti_2O_7$ [16] have revealed the amount of residual entropy is significantly reduced from the Pauling entropy in $Dy_2Ti_2O_7$ over much longer time scale, signaling disappearance of the spin-ice state and a first-order transition to the long-range ordered phase out of the spin-ice manifold, caused by the perturbations, viz., long-range nature of dipolar interactions beyond nearest neighbors and spin-phonon coupling due to lattice distortions [17] etc. $Dy_2Ge_2O_7$ [18-20] is arguably another DSI which lies more close to AF-DSI phase boundary [20]. On the other side, $Dy_2Zr_2O_7$ crystallizes into disordered fluorite structure and exhibits spin-liquid-like characteristics with complete disappearance of zero-point entropy [21].

The $A_2B_2O_7$ pyrochlore family can be extended to include the mixed pyrochlores, $A_2B'^{3+}B''^{5+}O_7$, in which $B'$ and $B''$ ions are distributed randomly at six-coordinated B sites [22], provided the ionic radius ratio ($r_A/r_B$) [23] lies within the range of 1.46–1.8 for the structurally ordered pyrochlores [22]. A renewed interest in the mixed pyrochlores emerged from the report of spin-ice-like state observed in $Dy_2ScNbO_7$ [12] with reduced zero-point entropy value than the canonical DSI, $Dy_2Ti_2O_7$. The cation disorder arising due to the mixed occupancy at B-site in $Dy_2NbScO_7$ could break the strict 'spin-ice' for the arrangements of the $Dy^{3+}$ spins leading to a partial release of the zero-point entropy [12]. On the other hand, in $Dy_2FeSbO_7$, spin-ice like configurations of $Dy^{3+}$ spins has been observed, excluding the effect of $Fe^{3+}$ moments, which order ferromagnetically, on the neighboring $Dy_4$ tetrahedra [24]. Similarly in another B-site diluted compound, $Dy_2Ti_{2-y}Fe_yO_7$, the spin dynamics are suppressed due to the Dy-Fe interactions and altered CF interactions [25]. The underlying principle in the backdrop of spin-ice physics is that the variations of lattice constant, ($r_A/r_B$) ratio, n.n. spin-spin (A-A) distance, etc. imposed by the chemical substitution of the B-site ions over any pyrochlore spin-ice series induce change in chemical pressure on the rare-earth network [20], which in effect leads to the variation of the relative strength and magnitude of the competing interactions over the spin-ice series. Another intriguing observation in the context of spin-ice dynamics is the emergence of magnetic (anti-) monopole-like excitations in, e.g., $Dy_2Ti_2O_7$ [2, 26-30], and $Dy_2Ge_2O_7$ [19], generated by the flipping of one spin out of "two-in/two-out" spin-ice configurations.

Given this background, the search for new spin-ices is, therefore, an active field of condensed-matter physics research and has drawn the attention of the broader science community. In this work, we have investigated the low-temperature properties and spin dynamics of a mixed pyrochlore, $Dy_2GaSbO_7$ (DGS) through the coordination of measurements of dc magnetization, ac susceptibility, and heat capacity, and CF computation, and found that DGS is a new spin-ice compound, but with increased zero-point entropy, unlike pure DSIs.

## II. EXPERIMENTAL

The polycrystalline $Dy_2GaSbO_7$ (DGS) sample were prepared by conventional solid-state reaction method using high-purity powders of $Dy_2O_3$, $Ga_2O_3$ and $Sb_2O_5$. All starting materials were preheated to 200°C for 6 hours to remove any moisture. The stoichiometric amounts of all precursor powders were then ground thoroughly, pelletized, and subsequently calcined for a total ~44 hours in a high-temperature muffle furnace in presence of air at 1320°C, with intermittent grinding and pelletizing. To identify the crystalline phase of the as-prepared powder samples, the X-ray powder diffraction (XRD) pattern was recorded (Fig. 1) at room temperature by scanning with a speed of 3° /min. over a range of $2\theta = 10°- 90°$ at 0.02° step-intervals using a Rigaku Miniflex 600 bench-top powder diffractometer which operates at 40 kV and 15 mA. The positions of the first peak (111) and the most intense peak (222) at (15.15°, 30.42°), respectively, are consistent with the pyrochlore structure and no impurity peaks were found. The structure analysis has been done by the Rietveld refinement method using Fullprof program on the XRD data. The refinement results indicated a good agreement between the observed and calculated Bragg's intensities of DGS (Fig. 1) for the pyrochlore-type structure (space group $Fd3m$). The occupancy ratio of Dy, Ga, Sb and O atoms are changed very nominally from their ideal values and found to be 0.99:0.48:0.52:1 in refinement. The lattice parameter $a_0$ and 48$f$ oxygen positional $x$-parameter were found to be 10.207(1) Å and 0.3275(2), respectively, in agreement with the literature [31]. The detailed structure analysis and characterization of DGS sample by optical spectroscopy were presented in our previous paper, Ref. [32].

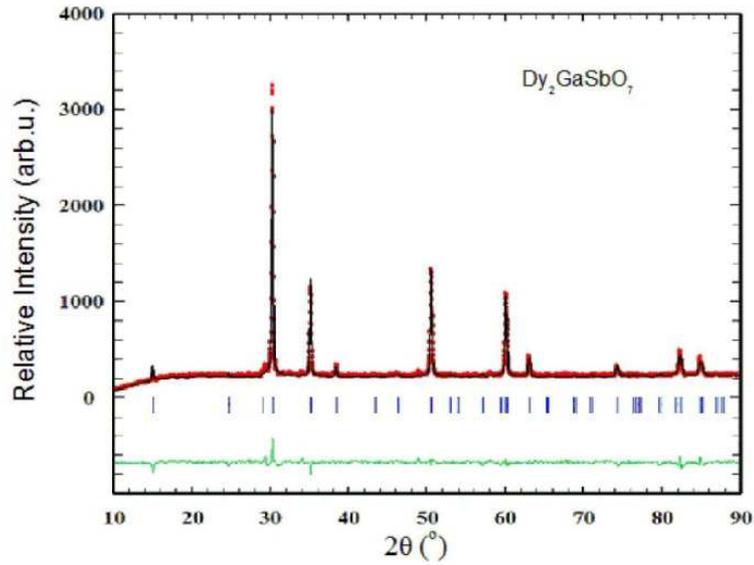

FIG. 1. XRD profiles of $Dy_2GaSbO_7$: observed (red points), calculated (solid lines), and the difference (green lines). Vertical marks (in blue) correspond to the allowed Bragg reflections for cubic space group $O_h^7 - Fd3m$. Most intense and significant ($hkl$) reflections are also denoted.

The magnetization $M(T,H)$ of the sample in the shape of a pellet (mass ~ 28.8 mg) was measured using a vibrating sample magnetometer attachment to physical property measurement system (PPMS: Quantum Design). Zero-field-cooled (ZFC) warming and field-cooled (FC) warming modes of $M(T)$ measurements were carried out between $T = 2-300$ K with applied magnetic field of $\mu_0H = 0.1$ T and 1 T. Isothermal magnetization $M(H)$ data at different temperatures, $T = 2, 5, 10, 20, 50$ and 100 K, were also recorded with varied fields up to $\mu_0 H = \pm 9$ T. The AC magnetic susceptibility $\chi_{ac}(f,T)$ was measured as a function of temperature within the range of 2 K– 40 K in the excitation frequency range $f$ = 110 Hz – 10 kHz using the ACMS option of the PPMS at ac driving field, $H_{ac}$ = 8 Oe.

Heat capacity $C_P$ of the pellet-shaped sample (mass ~ 8.2 mg) at constant pressure $p \approx 1.2$ mPa was measured at different fields, $H$ = 0-6T down to 0.39 K, employing a standard thermal-relaxation technique using the heat capacity attachment to the PPMS above 2 K and a $^3$He option for measurements below 2 K. The accuracies of heat capacities are within ± 2% over the range from 10 K to 270 K, ± 3% over the temperature range, $T$ = 2–10 K, and ± 2% below 2 K.

## III. RESULTS AND ANALYSIS

*DC magnetization*:

The temperature-dependent dc magnetic susceptibility, $\chi(T)$, of polycrystalline DGS for applied magnetic field of $\mu_0 H$ = 0.1 T and 1 T are plotted in Fig. 2 after demagnetization correction to the applied field using the expression, $\mu_0 H_{eff} = H - 4\pi N_d M$. The demagnetization factor, $N_d$, is taken to be 1/3 assuming spherical shape of the pellet sample [33]. The $\chi(T)$ data increase smoothly and do not show any distinct anomaly down to $T$ = 2 K, which is evidence of paramagnetism of Dy$^{3+}$ magnetic ions in DGS. The values of $\chi(T)$ for $H$ = 0.1 T almost coincide with the data for the field of 1 T in the temperature range of 300–20 K, implying that the temperature-dependent magnetization $M(T)$ increases linearly by ten times as the external field increases from 0.1 T to 1 T. Below $T$ = 20(±4) K, $\chi|_{1T}$ becomes lower in magnitude than $\chi|_{0.1T}$ (e.g., $\chi|_{0.1T} /\chi|_{1T}$ ~ 4.7 at $T$ = 2 K), implying the occurrence of a spin-spin correlation in the Dy tetrahedral network below $T \leq 20$ K at high field in competition with Zeeman interaction. Such magnetic feature was also observed in Dy$_2$Ge$_2$O$_7$ [18], and Dy$_2$Zr$_2$O$_7$ [34]. Thermal hysteresis is not observed in the ZFC and FC modes of susceptibilities in the range $T$ = 2–300 K, which rules out the formation of any magnetic cluster or spin-glass-like freezing in DGS.

The $\chi(T)$ data were fitted to a Curie-Weiss (CW) law, $\chi(T) = C/(T-\theta_{CW}) + \chi_0$, where $C = N\mu_{eff}^2/3k_B$ is the Curie constant, and $\chi_0$ (= $\chi_V +\chi_{dia}$) measures the contributions from temperature-independent Van Vleck paramagnetic ($\chi_V$) and inner-shell diamagnetic susceptibility ($\chi_{dia}$) to the measured dc susceptibility. The value of $\chi_{dia}$ of DGS is estimated to be $-0.71\times10^{-4}$ emu/mol-Dy by adding the contributions from Dy$^{3+}$, Ga$^{3+}$, Sb$^{5+}$ and O$^{2-}$ ions [35]. A slight downward curvature in the $\chi^{-1}(T)$ data is observed around 50 K, and therefore, shown in inset to Fig.2 , the CW fit was performed at two different temperature regimes [36]: (i) the fit of $\chi^{-1}(T)$ data for $H$ = 0.1 T within the high-temperature range of 300–100 K yields an effective magnetic moment $\mu_{eff}$ = 10.52(2) $\mu_B$/Dy and $\theta_{CW}$ = − 4.93(1) K. The negative $\theta_{CW}$ is indicative of net AF interactions among Dy spins at higher temperatures. (ii) on the other hand, the $\chi^{-1}(T)$ data when fitted over 2 K ≤ $T$ ≤ 50 K yields $\mu_{eff}$ = 10.0(1) $\mu_B$/Dy and $\theta_{CW}$ = +1.62(1) K. The positive value of $\theta_{CW}$ is a 'robust measure' [9] of effective FM interactions among the $Dy^{3+}$ spins developed at low-$T$(K). The value of $\mu_{eff}$/$Dy^{3+}$ ion extracted from the high-$T$(K) fit is close to the free-ion value of $\mu_{fi}^{Dy}$ = 10.646 $\mu_B$/Dy expected for the $^6H_{15/2}$ ground multiplet of $Dy^{3+}$-ion, and the low temperature value is equal to the value of $\mu$ = 10.0 $\mu_B$/Dy for a ground state with $J_z$ = ±15/2 double. Thus the ground doublet state plays a major role in tuning the magnetic properties at low temperature. The variation of $\mu_{eff}$ with temperature is attributed to the CF mixing up of the $J$ = 15/2 ground states with contributions from higher states within the $J$ manifold by intermediate-coupling mechanism. The value of $\chi_0 \approx \chi_V$ was found to be 0.002(1) emu/mol-Dy from the CW fit and matches with the value obtained for $Dy_2Ti_2O_7$ [33].

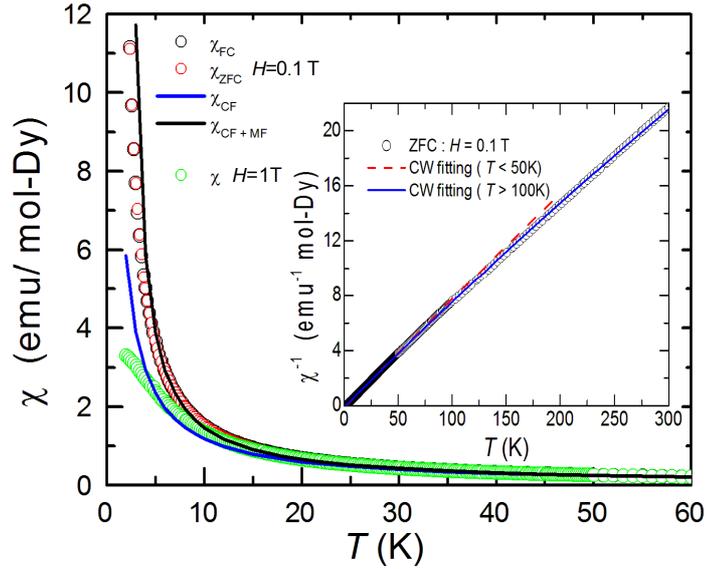

FIG.2. DC magnetic susceptibility, $\chi(T)$, of $Dy_2GaSbO_7$ measured in the magnetic field of $H$ = 0.1 T and 1 T with ZFC and FC mode. Straight lines are fit to $\chi(T)$ using CF theory (blue) and within the frame work of mean-field approximation (CF+MF) (black). Inset: CW fit to the inverse susceptibility.

To make a closer inspection on the variation of susceptibility data with temperature, we have examined the experimental $\chi T$ of DGS against 1/$T$. The $\chi T$ values for both fields coincide at high temperatures and exhibit a trough at 1/$T$ ≈ 0.02 $K^{-1}$ or $T$ = 50 K, above which the CF effect dominates and Ising approximation loses its validity thereafter [37]. The $\chi T$ data show a distinctly linear behavior for $\mu_0 H$ = 0.1 T and 1 T within lower temperature region of 20−2 K; for the lower field, $\chi T$ data increase linearly above $1/T \approx 0.05$ K$^{-1}$, while for the higher field, $\chi T$ decreases smoothly. Similar signatures are also obtained for Ho$_2$Ti$_2$O$_7$ and Yb$_2$Ti$_2$O$_7$, and Siddharthan *et al* [37] attributed such behavior to the competing effect and the relative strength of the FM long-ranged dipolar interactions and AF n.n. superexchange interactions among the magnetic Ising spins within tetrahedral network.

The isothermal magnetization $M(H)$ of Dy$_2$GaSbO$_7$ as a function of $H$ at some selected temperatures are drawn in Fig. 3. At $T = 2$ K, the $M(H)$ isotherm exhibits a steep increase below 1 T [31] and tends to attain thereafter a saturation value of $M_{sat} = 5.06\mu_B$/Dy above $H \approx 8$ T. This value is nearly 48% of the theoretical $\mu_{eff} = 10.646$ $\mu_B$/Dy value or half of the Brillouin value ($g_J J_z = 10$ $\mu_B$/Dy) of free Dy$^{3+}$ ion and is comparable to the $M_{sat}$ value obtained for other Dy-based pyrochlores Dy$_2$Ti$_2$O$_7$ [33], Dy$_{2-x}$Ca$_x$Ti$_2$O$_7$ [38], Dy$_2$Zr$_2$O$_7$ [21] and Dy$_2$Ge$_2$O$_7$ [18]. The magnetization isotherms can be explained with powder averaging the polycrystalline data by considering the effective Ising-type spin-1/2 ground state with transverse and longitudinal g-factors (g$_\perp$ = 0 and g$_\parallel$ = g$_z$), where Z-axis is along local <111> axes of the elementary tetrahedron. The magnetic moment per ion for this model at any temperature $T$ is given by [33],

$$M(H) = \frac{2}{g_z}\left(\frac{k_B T}{\mu_B H}\right)^2 \cdot \int_0^x x\tanh(x)dx + \chi_V H \qquad (1)$$

where, $x = g_z \mu_B H / 2k_B T$. Eq.(1) describes the measured data very well (solid curves in Fig.3) with $g_z$ = 19.3±0.2, which is close to the $g_z = 2g_J J_z = 20$, as expected for the pure Kramers doublet $J_z = \pm 15/2$ states of Dy$^{3+}$.

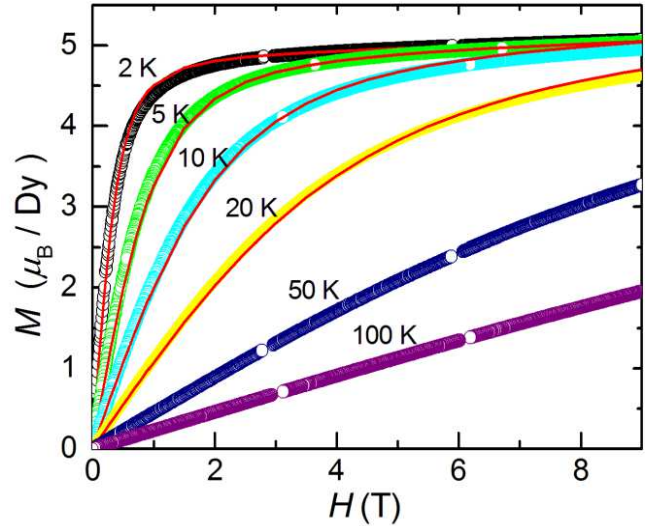

FIG. 3. Field dependencies of isothermal magnetization of Dy$_2$GaSbO$_7$ at several temperatures. Staright lines are fitting within Ising spin ½ approximation.

The $M^2$ vs $H_{eff}/M$ isotherms, namely Arrott plot, for DGS are analyzed using Arrott equation of state, $\mu_0 H_{eff}/M = a + bM^2$, where $a$ and $b$ are temperature-dependent coefficients. The value of $b$ which measures the slope of the curves is found to be positive ($b = +0.004$ T(ion/$\mu_B$)$^3$ at 2 K increasing to $+0.06$ T(ion/$\mu_B$)$^3$ at 100 K), in the overall magnetic measurements, implying that a second-order transition occurs at CW temperature, $T \sim O(\theta_{CW})$, in Dy$_2$GaSbO$_7$ in accordance with the Banerjee criterion [39].

*Crystal-field computation*

The dc magnetic susceptibility was calculated using a Van Vleck equation within the framework of single-ion crystal-field (CF) theory in combination with the molecular field (MF) [24,40]. In this model, the CF Hamiltonian ($H_{CF}$) is expressed by six CF interaction parameters (CFPs) in trigonal symmetry and the Zeeman interaction Hamiltonian contains spin-spin exchange tensors, e.g., $\lambda_\parallel$ and ($\lambda_{\perp,x}$ and $\lambda_{\perp,y}$), which are longitudinal and transverse to the R-R bond direction. Calculated susceptibility with CF+MF model was plotted in Fig.2. Detailed description of the computation was given in Appendix A. We found that the ground CF state, $|\Psi_g\rangle$, is a well-isolated by an energy separation, $\Delta_{CF} \approx 192$ K, from the first excited doublet and has major contribution from the $J_z = \pm 15/2$ state with $g_\parallel = 19.1(1)$ and $g_\perp = 0$. The low-temperature magnetic properties of the material can, therefore, be modeled by an effective pseudo-spin $\tilde{S} = 1/2$ system [1,3,10], oriented along local <111> axes.

The interaction Hamiltonian between the n.n. Ising-spins ($\sigma_i = \pm 1$ at site $i$) can, therefore, be expressed by the following form, $H_{eff} = J_{eff}^{(1)} \sigma_1 \sigma_2$ [9,10], where $J_{eff}^{(1)} = J_{nn} + D_{nn}$ is defined as the effective energy scale of n.n. interactions [9,10]; $J_{nn} = (2\lambda_\parallel - \lambda_{\perp,x})(g_\parallel \mu_B/2)^2/3$ [40] and $D_{nn} = (5/3)(g_\parallel \mu_B/2)^2/r_{nn}^3$ are, respectively, the effective n.n. exchange and dipolar coupling [10, 40]. Using the values of the parameters derived above, we obtained $J_{nn} = -0.752(1)$ K and $D_{nn} = +2.01$ K at low temperatures, and hence $J_{eff}^{(1)} = +1.26(1)$ K. Thus though the n.n. exchange interactions between Dy$^{3+}$-spins is AF, due to dominance of the FM dipolar interactions in the n.n. scale, $J_{eff}^{(1)} > 0$ signifying effective FM interactions for n.n. spin-ice criterion [9,10] and giving rise to a positive $\theta_{CW}$ in DGS. Typically the frustrated pyrochlores in which n.n. dipolar interactions are the leading energy interactions (due to large $\mu_{eff}$) with n.n. FM coupling $J_{eff}^{(1)}$ and a critical range of the ratio $J_{nn}/D_{nn} > -0.91$ and $J_{eff}^{(1)}/D_{nn} > 0.095$ [9, 15] are referred to as 'standard dipolar spin-ice' (s-DSI) [10,15,41]. In the present case, $J_{nn}/D_{nn} = -0.374$ in comparison to the values, $-0.46$, $-0.5$, $-0.56$, respectively, for Dy$_2$Sn$_2$O$_7$, Dy$_2$Ti$_2$O$_7$ [20] and Dy$_2$Pt$_2$O$_7$, [42] and $-0.27$, $-0.35$ for Ho$_2$Ti$_2$O$_7$ and Ho$_2$Ge$_2$O$_7$ [20], which reveals that DGS is more inside the DSI region of the magnetic phase boundary than its Dy counterparts [20]. Table I enlists some selected structural and magnetic parameters of these DSIs. The variation in the relative strength of magnetic interactions through any set of pyrochlore DSI series is attributed to 'chemical pressure effect' due to B-ion substitution [20]. Using the above values of $J_{nn}$ and $D_{nn}$, one could obtain a rough estimate of the second and third n.n. interactions, in terms of $J_{eff}^{(2)} = D_{nn}/(15\sqrt{3}J_{nn})$, and $J_{eff}^{(3)} = D_{nn}/(40J_{nn})$ [17] to consider the long-range nature of exchange interactions (g-DSI model) [41], taking account of the effect of distortions in the lattice sites which causes the local changes in magnetic correlations and are referred to as d-DSI model [17]. The values are $J_{eff}^{(2)} = -0.103$ K and $J_{eff}^{(3)} = -0.067$ K, which are antiferromagnetic and much weaker than $J_{eff}^{(1)}$. An estimate of the dipolar and exchange contributions to the CW temperature can also be obtained from the relations $\theta_{dip} = 16\pi(g_{\parallel}\mu_B)^2(3.7607 - N_d)/9a_0^3$ and $\theta_{ex} = (2\lambda_{\parallel} - \lambda_{\perp,x})(g_{\parallel}\mu_B)^2/6$ [40] and found to be, respectively, +3.29 K and −1.51 K. Hence in the low-temperature zone (where $\theta_{CF} \approx 0$ [24]), the value of $\theta_{CW} = \theta_{dip} + \theta_{ex} = +1.78$ K which agrees with the value ($\theta_{CW} \approx +1.62$ K) obtained from the CW analysis of the low-temperature susceptibility.

TABLE I. Lattice parameters and selected magnetic parameters for Dy-based pyrochlore spin-ices.

| | $a_0$ (Å) | $r_A/r_B$ | $x$ | $\theta_{CW}$ (K) | $J_{nn}$ (K) | $D_{nn}$ (K) | $J_{eff}^{(1)}$ (K) | $J_{nn}/D_{nn}$ | $J_{eff}^{(2)}$ (K) | $J_{eff}^{(3)}$ (K) | $T_{si}$ (K)[c] |
|---|---|---|---|---|---|---|---|---|---|---|---|
| Dy$_2$Sn$_2$O$_7$ [20] | 10.4 | 1.488 | 0.337 | +1.7 | −1.0 | 2.15 | +1.16 | −0.46 | | | 5.2 [12] |
| Dy$_2$Pt$_2$O$_7$ [43] | 10.19 | 1.643 | | +0.77 | −1.28 | 2.29 | 1.01 | −0.56 | | | |
| Dy$_2$Ti$_2$O$_7$ [20] | 10.124 | 1.697 | 0.323 | 1.1 | −1.15 | 2.35 | +1.20 | −0.49 | −0.14[a] | 0.025[a] | 4.5 [61] |
| Dy$_2$Ge$_2$O$_7$ [20] | 9.93 | 1.938 | | 0 | −1.83 | 2.5 | +0.67 | −0.73 | | | |
| Dy$_2$FeSbO$_7$ [24] | 10.261 | 1.65 | 0.33 | +2.42 | −0.313 | 1.93 | +1.62 | −0.16 | −0.237 | −0.154 | 12.1 |
| Dy$_2$GaSbO$_7$ | 10.207 | 1.684 | 0.327 | +1.62 | −0.752 | +2.01 | +1.26 | −0.374 −0.383[b] | −0.103 | −0.067 | 9.57 |

[a] Taken from [41].

[b] The value of $J_{nn}/D_{nn}$, obtained from the variation of $T_{peak}$, at which heat capacity peak appears, against $J_{nn}/D_{nn}$ [15] for DSI model, is found −0.383 at $T_{peak}/D_{nn} \approx 0.676$ for DGS and hence $J_{nn} = -0.770$ K.

[c] $T_{si}$ is the temperature at which integrated magnetic entropy reaches to spin-ice value of 4.1 J K$^{-1}$(mol-Dy)$^{-1}$ when the field of $H = 1$T is applied. $T_{si} \approx 7.5$ K for Dy$_2$NbScO$_7$ [12].

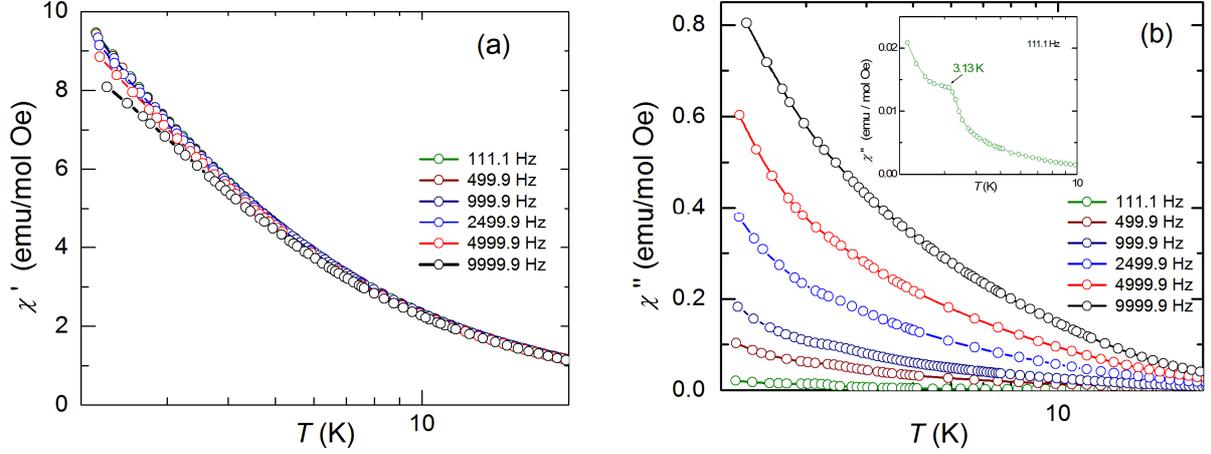

FIG.4. Real $\chi'$ (a) and imaginary $\chi''$ (b) parts of the ac susceptibility $\chi_{ac}(T)$ vs temperature for $Dy_2GaSbO_7$ sample spanning the frequency range $f$ = 111–10000 Hz. Inset of (b) shows spin-freezing anomaly at $T_{ice}$ ~ 3.13 K at $f$ = 111 Hz.

*AC magnetic susceptibility*:

To investigate the spin-ice dynamics, the temperature-dependent ac susceptibility $\chi_{ac}(T)$ data of DGS were plotted in Fig. 4 at different frequencies between 111 Hz – 10 kHz in the temperature range 2 K to 40 K. The real part ($\chi'$) of $\chi_{ac}(T)$ increases as temperature decreases and does not show any anomaly within our measurement frequency window (Fig. 4a) down to 2 K, similar to $\chi_{dc}$. The $\chi'$ is independent of frequency down to $T$ ~ 8 K, below which frequency dependence is clearly developed for higher frequencies and $\chi'(T)$ gradually reduces. Unlike paramagnetic-like behavior of $\chi'(T)$, frequency dependency becomes more pronounced in the imaginary part $\chi''(T)$ of $\chi_{ac}(T)$ which increases differently with lowering temperature as frequency increases. A very small spin-freezing anomaly develops at lower temperature, (e.g. $T_{ice}$ ~ 3.13 K at $f$ = 111 Hz), shown in inset to Fig.4(b), which shifts towards higher temperatures (e.g., at 3.32 K for 500 Hz, at 3.43 K for 1 kHz) with increasing frequency and may disappear at higher frequency, $f \geq 2.5$ kHz. In contrast to DGS, it may be recalled that in canonical Dy-based DSIs, e.g., $Dy_2Ti_2O_7$ (DT) [43-45] and $Dy_2Sn_2O_7$ (DS) [12, 46], two dynamic spin freezings are observed in $\chi_{ac}(T)$ curves at high temperature (at ~16 K for DT, ~25 K for DS), followed by a lower-$T$ peak (at ~2-4K for DT and DS). The high-$T$ peak corresponds to occurrence of spin-spin correlations in short-range scale [45] or single spin freezing [47], while low-$T$ freezing is associated with the formation of spin-ice state [48,49]. The frequency dependence of the spin-freezing temperature, $T_{ice}$, of DGS is roughly fitted to an Arrhenius formula, $f = f_0 \exp(-E_a/T_{ice})$, where $f_0$ is characteristic frequency, and $E_a$ is activation energy for spin relaxation [12,43]. The value of $E_a$ is found to be 126±4 K which is lower than the obtained values for $Dy_2Ge_2O_7$ (162 K) [18], $Dy_2FeSbO_7$ (~165 K) [24], $Dy_2Ti_2O_7$ (~200 K) [43], $Dy_2Sn_2O_7$ ($\approx$ 220 K) [12], and $Dy_2Pt_2O_7$ (~267 K) [42]. In contrast, much lower value of $E_a$ was reported for a mixed pyrochlore compound, e.g., 42 K for $Dy_2Sn_{1.75}Sb_{0.25}O_{7.125}$ [12]. The spin relaxation mechanism and the energy barrier $E_a$ are associated with the crystal-field perturbations [43], and therefore, cation disorder from the random occupancy at B site elements may cause the lower value of $E_a$ in the mixed pyrochlores [12].

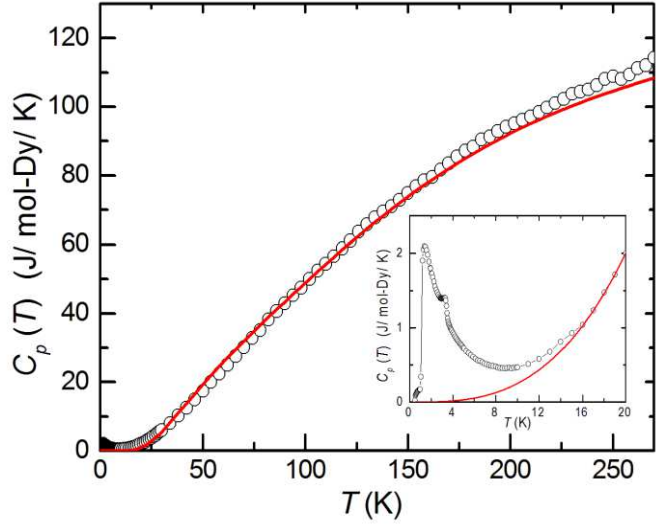

FIG.5. Heat capacity $C_p$ versus $T$(K) for $Dy_2GaSbO_7$ measured at $H = 0$ T (symbol). Solid lines represents the calculated values considering the contributions from Debye, Einstein and CF (Schottky) heat capacities (see Appendix B). Inset shows the lattice heat capacity using Debye $T^3$ formula (line) below 20K.

*Heat capacity*:

The total heat capacity $C_p(T)$ of DGS has been displayed in Fig. 5 for $H = 0$ T between the temperature range of 0.35 K and 270 K. The field-dependence of $C_p(T)$ are also plotted in Fig.6 between 0.35 K and 30 K for the external magnetic field of $H = 0, 1, 2, 3, 6$ T. The zero-field heat capacity shows a rounded peak at $T_{peak} \approx 1.36(2)$ K, which is of the order of $J_{eff}^{(1)}$, as were reported for other Dy-based dipolar spin-ice pyrochlores, e.g., at 1.2 K for $Dy_2Sn_2O_7$ [20], at 1.13 K for $Dy_2Pt_2O_7$ [42], at 1.25 K for $Dy_2Ti_2O_7$ [11,20], at ~1 K for $Dy_2Sn_{2-x}Sb_xO_{7+x/2}$ [12]. Thus one may notice that the peak position $T_{peak}$ shifts to higher temperature as $r_A/r_B$ ratio increases for the Dy series (Table I), though the dependence is not linear. The peak amplitude, $C_{peak} \approx 2.1$ J K$^{-1}$(mol-Dy)$^{-1}$, is considerably lower than the mean-field prediction of 12.5 J K$^{-1}$(mol-Dy)$^{-1}$ expected for a long-range second-order magnetic transition for the effective spin $\tilde{S} = 1/2$ system [50]. The broad maximum and the absence of any sharp feature in $C_p(T)$ is a strong indication that the system does not develop any long-range magnetic order via a thermodynamic phase transition [9]. The $T_{peak}$ shifts to the higher temperatures and broadens as applied magnetic field increases, and finally smears out at higher field. In addition, shown in inset to Fig.5, a kink at $T \sim 3.25(2)$ K is also observed which can be identified with the spin-freezing $T_{ice}$ observed in the ac susceptibility $\chi''(T)$ and vanishes when field is applied.

We extract the magnetic specific heat, $C_{mag}$, by subtracting the ($C_{lat} + C_{Sch}$) from the measured $C_p(T)$ data, and $C_{mag}/T$ vs $T$ of DGS are drawn in inset to inset (a) to Fig.6. The estimation of lattice ($C_{lat}$) and Schottky ($C_{Sch}$) components of heat capacity is illustrated in Appendix B. In addition to two anomalies at 1.32(2) K and 3.28(3) K, the $C_{mag}/T$ at zero-field exhibits an upturn at 0.95(1)K, followed by a round-shaped very low peak at 0.69(1) K, which disappears at $H$ =1 T. Interestingly, similar upturn was also reported for $Dy_2Ti_2O_7$ at $T \le 0.6$ K [16,17], though the peak height is much smaller for DGS. This peak is not associated with nuclear hyperfine heat capacity, which peaks at 20 mK and extends up to $T \sim 0.4$ K, as found for $Dy_2Ti_2O_7$ [51,52]. The random disorder, either in the form of low level of stuffing of rare-earth ions (A site) onto the B-sites [53-55], or oxygen vacancies [56], disorder in the B site ions in mixed pyrochlores [12], may cause the small distortions in the lattice connectivity between neighbors [17]. Such random disorder may, by lowering the local symmetry of the crystal-field, destroy the local Ising nature of the moments and induce quantum fluctuations within the Dy-tetrahedra, and hence is the most likely source of the upturn [52].

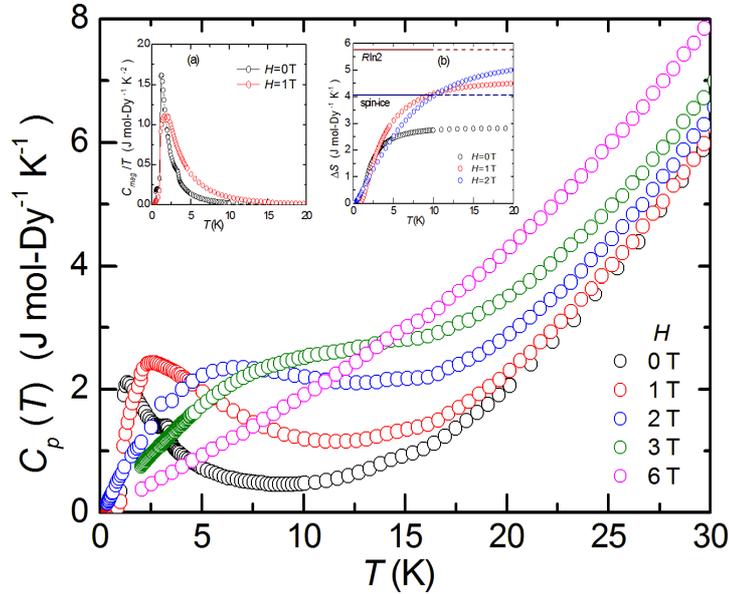

FIG.6. Heat capacity $C_p$ against $T$(K) for $Dy_2GaSbO_7$ at different magnetic fields (symbol). Inset: (a) the $C_{mag}/T$ at $H$ = 0,1 T; (b) integrated values of magnetic entropy versus temperature in $H$ = 0, 1, 2 T.

The $C_{mag}/T$ curve shifts to the higher temperature as field increases, indicating that the area under the curve (i.e., entropy) increases with the field. The magnetic entropy change $\Delta S_{mag}$ for $H = 0$, 1, and 2 T are calculated (inset b to Fig.6) by integrating $C_{mag}/T$ within the temperature range, $T \leq 20$ K, $\Delta S_{mag} = \int_{T_0}^{T}(C_{mag}/T)dT$, by crudely extrapolating $C_{mag}/T$ to its zero value. We found that the integrated entropy increases with increasing temperature up to 3.25 K ($\sim T_{ice}$) and finally saturates at $T \geq 10$ K to the value of $\Delta S_{sat} \approx 2.82(1)$ J K$^{-1}$(mol-Dy)$^{-1}$. The difference between the $\Delta S_{sat}$ and the expected value of $S = R \ln 2 = 5.76$ J K$^{-1}$ mol$^{-1}$ corresponding to the full $2N$ states available to $N$ Ising $\tilde{S} = \frac{1}{2}$ spins system in the high-temperature regime ($T \gg \theta_{CW}$) [9] is the *zero-point entropy* ($S_0$), and this difference constitutes the primary thermodynamic evidence for the existence of a spin-ice state. Hence the calculated zero-point entropy of DGS for $H = 0$ T is 2.94 J K$^{-1}$(mol-Dy)$^{-1}$, which is much larger than the Pauling spin-ice value of $S_0 = 1.68$ J K$^{-1}$ mol$^{-1}$, indicates additional degeneracy in DGS, which may be induced by disorder, compared to the macroscopic degeneracy ($\Omega_0 = 2^N (6/16)^{N/2}$) in s-DSI system. We hypothesize that out of possible $2^4$ spin configurations on a tetrahedron, if four configurations remains thermally inactive or energetically constrained due to disorder, and six obeys the conventional 'two-in/two-out' spin-ice rule per tetrahedron, then following Pauling-like argument [9,47], we could obtain $S_0 = k_B \ln[2^N (6/12)^{N/2}] \approx 2.88$ J K$^{-1}$ mol$^{-1}$, which agrees with the estimated zero-point entropy for DGS. Application of a magnetic field increases the Zeeman energy of the spins, thereby effectively lowering the energy barriers for spin orientation among the different manifolds of the ground state spin configuration and lifts the degeneracy of the ground state [11], which restores long-range order to gain some entropy [53]. We note that the spin-ice entropy ($\sim 4.1$ J K$^{-1}$ mol-Dy$^{-1}$) is recovered at the temperature $T_{si} = 9.57(1)$ K on applying a field of 1T, with partial release ($\sim 41\%$) of residual entropy, and attains the saturation value of 4.50(1) J K$^{-1}$(mol-Dy)$^{-1}$ and 5.0 J K$^{-1}$(mol-Dy)$^{-1}$ at 20 K for $H = 1$T and 2T, respectively (Fig.7). That is, the saturation entropy is 22% and 13% smaller than the value of $R\ln 2$ at $H = 1$ T and 2T, respectively, in contrary to the pure pyrochlore DSIs, e.g., $Dy_2Ti_2O_7$, $Dy_2Sn_2O_7$ and $Ho_2Ti_2O_7$, for which the saturation entropy at $H = 1$T is 15% less than $R\ln 2$ [11,12,14,57]. In other words, high magnetic field is required for DGS to make it entropic equivalent to the pure DSIs. Further, in Table I, we note that the values of $T_{si}$ for the mixed pyrochlores, $Dy_2NbScO_7$, $Dy_2FeSbO_7$, and $Dy_2GaSbO_7$, are greater than for, e.g., $Dy_2Ti_2O_7$ and $Dy_2Sn_2O_7$, which reveals that the thermal barrier or constraint for spin reorientation is stronger in mixed pyrochlores than pure DSIs.

## IV. DISCUSSION AND CONCLUSION

The dc and ac magnetic susceptibility, and heat capacity with and without magnetic field have been investigated on a B-site mixed pyrochlore $Dy_2GaSbO_7$ to explore its spin dynamics. The salient observations may be summarized below:

(i) Dy magnetic moments in DGS behave as an ensemble of Ising $\tilde{S}$ = ½ spins in the ground state. The spins are oriented along the local <111> axes of each elementary tetrahedron due to strong single-ion crystal-field anisotropy in coordination with ferromagnetic dipolar and anti-ferromagnetic exchange interactions. The dipolar interactions are strong enough to counterbalance an AF exchange for DGS.

(ii) The spins freeze at temperature $T_{ice}$ ≈ 3.1 K corresponding to dynamical spin-freezing into a disordered state in compliance with the spin-ice rules, like cannonical dipolar spin-ices. However unlike DSIs, no single-spin freezing is observed for DGS. The energy barrier, $E_a$ ≈ 126 K, (which determines the tunneling barrier of the spin states through quantum and/or thermal processes [45,47]), associated with the spin relaxation mechanism, is much less than the energy separation ($\Delta_{CF}$ ≈192 K) of the first excited CF level from the ground state doublet. That is, the spin freezing transition in DGS is mainly controlled by the quantum mechanical tunneling [58] constrained by the inherent cation disorder at B sites.

(iii) The spin ensemble has a finite zero-point entropy amounting to 2.94 J K$^{-1}$(mol-Dy)$^{-1}$, which is much larger than the Pauling residual entropy for common 'water ice' or cannonical DSI pyrochlores. It is relevant to mention that the zero-point entropy depends nonmonotonically on dilution of Dy-spins at A-site by nonmagnetic ions and approaches to the value of $R$ln2 for free spins in the limit of high dilution [45,47], as well as on stuffing of A-site ions onto the B-sites. Further, the thermal constraint for spin reorientation is much stronger for the mixed pyrochlores than the pure pyrochlore DSIs. Since the B′ and B″ ions have different ionic radius in mixed pyrochlores, due to random occupancy of B-site ions, the equality of n.n. distance, A-A = B-B, is lost in the mixed pyrochlores (similarly for the stuffed DSI materials [17]) which induces the local changes in the magnetic interactions beyond nearest-neighbors as well as causes variation of chemical pressure on spin system. We argue that these effects have significant impact on the low-temperature magnetic and heat capacity properties of the present system.

(iv) The lattice constant, magnetic moment, and $J_{eff}$, which define the monopole contact distance and elementary monopole charge [19], of DGS are nearly identical for $Dy_2Ti_2O_7$, and $Ho_2Ti_2O_7$. Therefore, DGS is proposed as another ideal system to study monopole-antimonopole-pair excitations through the study of broadband spectroscopic measurements of dielectric function and field-dependent thermal conductivity. In this context, it would be very interesting to elucidate how these excitations are affected by the large thermal constraint and disorder in DGS.

To conclude, $Dy_2GaSbO_7$ is a spin-ice compound with complex dipolar and exchange interactions and having enhanced zero-point entropy, in variance with pure DSIs, due to random disorder and chemical pressure of B-site ions.

**ACKNOWLEDGEMENTS**

This work was supported by SERB funded project (file no. CRG/2018/000171) and partially by UGC-DAE-CSR (Indore) project (No. CSR-IC-249/2017-18/1330). The authors also acknowledge the facilities available through DST-PURSE & DST-FIST program of Dept. of Physics, University of Kalyani. Authors are also thankful to UGC-DAE-CSR, Kolkata center, for magnetic measurements.

**Appendix A**

**Combined crystal-field calculation:**

Single-ion crystal-field (CF) theory in coordination to the molecular field (MF) [40] has proved to be an efficient treatment to describe the temperature dependency of magnetization data and the nature of ground state of rare-earth ions depending on its local symmetry in the crystal. The CF at the $R^{3+}$-ion sites in the pyrochlores has trigonal point symmetry $\bar{3}m$ ($D_{3d}$), with the local $\bar{3}$ axes parallel to the <111>-type directions of the crystal lattice. The effective single-ion Hamiltonian has the form [40]:

$$H_t(i) = H_{FI}(i) + H_{CF}(i) + H_Z^{ext} + H_Z^{int}. \tag{A1}$$

$H_{FI}$ is free-ion (FI) Hamiltonian that operates in the 520 $|SLJM_J\rangle$ intermediate-coupled states of the $4f^9$ electronic configuration of $Dy^{3+}$-ions [24].

The CF Hamiltonian ($H_{CF}$) in the local system of coordinates with the Z-axis // [111] trigonal symmetry axis at the $Dy^{3+}$ site is expressed as,

$$H_{CF} = B_{20}U_{20} + B_{40}U_{40} + B_{60}U_{60} + B_{43}(U_{43} - U_{4-3}) + B_{63}(U_{63} - U_{6-3}) + B_{66}(U_{66} + U_{6-6}), \tag{A2}$$

where $U$'s are the one-electron intra-configuration unit tensor operators and $B_{kq}$'s are even parity CF interaction parameters (CFPs) [59,60]. The last two terms in Eq. (A1) represent the perturbation energy due to the Zeeman coupling, e.g., (i) between the R-moment and the effective magnetic field $\vec{H}_{eff}$, $H_Z^{ext} = -\vec{\mu} \cdot \vec{H}_{eff}$, and (ii) between the R-ion and the local internal magnetic field at the R-site appearing due to long-range magnetic dipole-dipole and anisotropic exchange interactions, $H_Z^{int} = \vec{\mu} \cdot \hat{H}_{int} = H_{dip} + H_{exch}$, described in Ref. [40]. The dipolar terms at R-sites $j$ are given by $H_{dip,j} = \sum_{j'} Q(j,j')\langle\mu(j')\rangle$, where the components of the tensor $Q$ are the corresponding dipole lattice sums which have been computed in the crystallographic system of coordinates in Ref. [40]. The spin-spin exchange tensor has three principal independent components which couple the n.n. spins along ($\lambda_{\parallel}$) and normal ($\lambda_{\perp,x}$ and $\lambda_{\perp,y}$) to the R-R bond direction. The values of the exchange coupling constants, $\lambda_{\parallel}$, $\lambda_{\perp,x}$ and $\lambda_{\perp,y}$, and six CFPs, were obtained from the fit to the dc susceptibility using a Van Vleck equation,

$$\chi_j^s = \frac{N_a\mu_B^2}{Z}\sum_{n,m}\{\frac{(E_{n,m}^{(1)})^2}{k_BT} - 2E_{n,m}^{(2)}\}\exp(-\frac{E_{n,m}^{(0)}}{k_BT}).$$ (A3)

In order to fit the susceptibility data, the starting CFPs, e.g., $B_{20} = -544.4$, $B_{40} = 2438.5$, $B_{60} = -687.2$, $B_{43} = 1392.2$, $B_{63} = -795.4$, $B_{66} = -1360.0$ (in cm$^{-1}$), in Eq.(A2) were taken from the inelastic neutron scattering results of Dy$_2$Ti$_2$O$_7$ [61] and varied these initial parameters first individually and then collectively in a self-consistent manner. The CFPs, $B_{20} = -395.0$, $B_{40} = 1840.0$, $B_{60} = -565.0$, $B_{43} = 650.0$, $B_{63} = 670.0$, $B_{66} = -1190.0$ (in cm$^{-1}$) are finally obtained from the best-fit to the susceptibility. The ±1% variation in CFPs varies the calculated susceptibility by only ±0.03% and the energy levels by ±0.6%. The calculated energies of the eight doublet sublevels arising from the splitting of $^6H_{15/2}$ ground multiplet of Dy$^{3+}$-ion are 0, 191.6, 307.0, 323.8, 400.0, 649.0, 744.5, 751.8 (all in K).

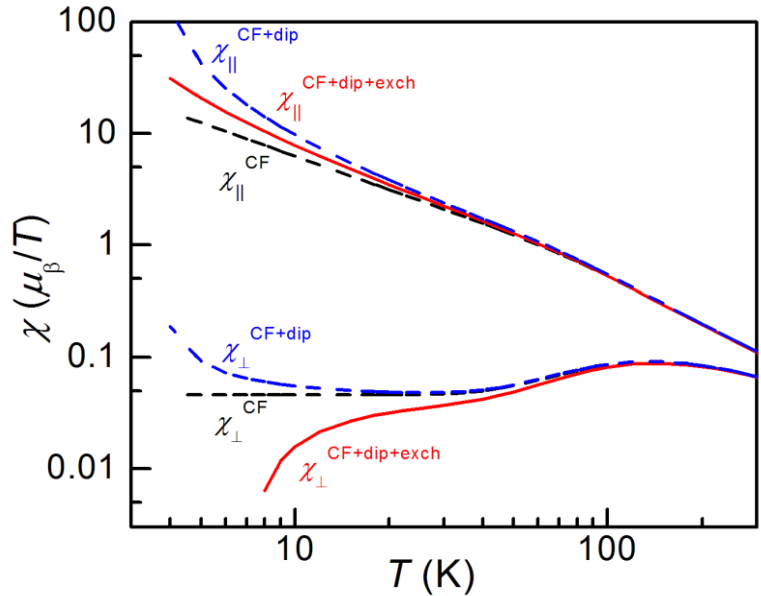

FIG. A1. The contributions of CF, dipolar and n.n. exchange interactions to the longitudinal ($\chi_{\parallel}$) and transverse ($\chi_{\perp}$) components of dc magnetic susceptibility.

The wave function of the ground CF doublet is approximated by $|\Psi_g\rangle \approx \mp 0.94|\pm15/2\rangle + 0.2|\pm9/2\rangle \pm 0.1|\pm3/2\rangle$, with $\langle\Psi_g|J_Z|\Psi_g\rangle = 7.16(1)$. Values of the Landé g-factors of $|\Psi_g\rangle$ are highly anisotropic, $g_\parallel = 19.1(1)$ and $g_\perp = 0$, which match with the values obtained in $M(H)$ calculation. The calculated magnetic moment of $Dy^{3+}$ ion in the ground state is $\mu_g = \langle g_J J_z\rangle = 9.55(1)\mu_B$/Dy which gives the ground state contribution to the effective magnetic moment derived from the CW fit of the susceptibility. The low-temperature magnetic properties of the DGS can, therefore, be approximated by an effective pseudo-spin $\tilde{S} = 1/2$ system with a large magnetic moment.

Calculated single-ion CF susceptibility, $\chi_{CF}(T)$, (and also $1/\chi_{CF}$) shows excellent agreement with the experiment (within 1.5% at 60 K), in the temperature range 60 K $< T <$ 300 K (Fig. 2), but decreases remarkably below 40 K (by 6% and 36% at 40K and 5 K, respectively). The dipolar and AF exchange interaction with coupling constants, $\lambda_\parallel = -0.039(1)$ T/$\mu_B$, $\lambda_{\perp,x} = -0.042(1)$ T/$\mu_B$, and $\lambda_{\perp,y} = 0$, were, therefore, introduced to describe the experimental $\chi(T)$; the calculated values (CF+MF) match very well (within 2-4% between 40–5K) with the experimental data down to 5±1 K. The components of the effective site susceptibility tensors along and normal to <111> axes, $\chi_\parallel$ and $\chi_\perp$, and the contributions of CF, dipolar and n.n. exchange interactions to $\chi$'s are displayed in Fig. A1. The values of $\chi_\parallel$ are found larger than $\chi_\perp$, which imply that the magnetic anisotropy at Dy site is Ising-like along the local <111> axes of the tetrahedron due to the strong CF axial anisotropy combined with spin-spin exchange and dipolar interactions among Dy-spins.

**Appendix B**

**Calculation for lattice and Schottky heat capacity:**

The $C_p(T)$ of $Dy_2GaSbO_7$ can be expressed as $C_p(T) = C_{lat} + C_{Sch} + C_{mag}$ within the temperature range studied here, where the components are, respectively, lattice specific heat ($C_{lat}$), Schottky anomalies ($C_{Sch}$) and magnetic specific heat ($C_{mag}$). In the low-$T$(K) zone, the lattice contribution is composed of the lattice vibrations (phonons) at constant volume ($C_v \approx C_p$) and electronic part, and is usually expressed by the Debye model [62] as, $C_{lat}(T) = \gamma T + \beta T^3$, where, $\beta$ and $\gamma$ are coefficients to lattice and the electronic contributions to the heat capacity, respectively. The value of $\beta$ is related to the Debye temperature, $\theta_D = \left(\frac{12\pi^4 nR}{5\beta}\right)^{\frac{1}{3}}$, where $n = 11$ is the number of atoms per f.u. The $C_p$ data of DGS in $T < 20$ K are fit to Debye model, displayed in inset to Fig.5, and found $\gamma = 0$ implying insulating character of DGS and $\beta = 5 \times 10^{-4}$ JK$^{-4}$/mol-Dy. Using the value of $\beta$, the value of $\theta_D$ of DGS was estimated to be 350(±2) K, which may be compared with the reported value for $Dy_2Ti_2O_7$ ($\theta_D$ = 283 K below 30 K) [63] and $Dy_{1.8}Ca_{0.2}Ti_2O_7$ ($\theta_D$ = 295 K below 19 K) [38]. Since the above $\theta_D$ value could not fit the heat capacity in high-$T$ regime due to presence of optic modes of lattice vibrations, $C_p(T)$ data were analyzed using a combination of Debye model (Eq.B1) and Einstein model (Eq.B2) of heat capacity.

$$C_{ph}^{Debye} = 9nR(T/\theta_D)^3 \int_0^{\theta_D/T} \frac{x^4 e^x}{(e^x - 1)^2} dx \tag{B1}$$

and

$$C_{ph}^{Einstein} = 3nR(\theta_E/T)^2 \frac{\exp(\theta_E/T)}{[\exp(\theta_E/T) - 1]^2} \tag{B2}$$

where $\theta_E$ is Einstein temperature. In this combined model, the $C_p(T)$ can be expressed as [38],

$$C_p(T) = wC_V^{Debye} + (1-w)C_V^{Einstein} + C_{Sch}, \tag{B3}$$

where $w$ is the fractional weightage of the Debye heat capacity in the total lattice heat capacity [38,64]. The Schottky contribution $C_{Sch}$ was evaluated by considering the thermal distribution of $Dy^{3+}$-ions over the obtained CF level scheme of the ground multiplet. The best fit to the thermal variation of ($C_p - C_{Sch}$) data is obtained when $w$ ~74%. The values of $\theta_D$ = 732(±3) K and $\theta_E$ = 156(±2) K were obtained from fit. The calculated $C_p(T)$ using Eq.(B3) for DGS at $H$ = 0 T is shown by solid red curve in Fig. 5 and fit reasonably well with the experimental $C_p(T)$ data above 20 K.

# References

[1] J.S. Gardner, M.J.P. Gingras, and J.E. Greedan, Rev. Mod. Phys. 82, 53 (2010).

[2] L. Balents, Nature 464, 199 (2010).

[3] S.T. Bramwell and M.J. Harris, J. Phys.: Condens. Matter 32, 374010 (2020).

[4] M.J.P. Gingras and P.A. McClarty, Rep. Prog. Phys. 77, 056501 (2014).

[5] S.-H. Lee, C. Broholm, W. Ratcliff, G. Gasparovic, Q. Huang, T. H. Kim, and S.-W. Cheong, Nature 418, 856 (2002).

[6] J.D.M. Champion, M.J. Harris, P.C.W. Holdsworth, A.S. Wills, G. Balakrishnan, S.T. Bramwell, E. Čižmár, T. Fennell, J.S. Gardner, J. Lago, D.F. McMorrow, M. Orendáĉ, A. Orendáĉová, D. McK. Paul, R.I. Smith, M. T.F. Telling, and A. Wildes, Phys. Rev. B 68, 020401 (R) (2003).

[7] P. Schiffer, Nature 420, 35 (2002); I. Mirebeau, I.N. Goncharenko, P. Cadavez-Peres, S.T. Bramwell, M.J.P. Gingras, and J.S. Gardner, Nature 420, 54 (2002).

[8] R. Sibille, E. Lhotel, M. C. Hatnean, G. J. Nilsen, G. Ehlers, A. Cervellino, E. Ressouche, M.Frontzek, O. Zaharko, V. Pomjakushin, U. Stuhr, H. C. Walker, D. T. Adroja, H. Luetkens, C. Baines, A. Amato, G. Balakrishnan, T. Fennell, and M. Kenzelmann, Nature Commun. 8, 892 (2017).

[9] S.T. Bramwell, M.J.P. Gingras, and P.C.W. Holdsworth, in *Frustrated Spin Systems*, edited by H. T. Diep (World Scientific, 2004), Chap. 7.

[10] S.T. Bramwell and M.J.P. Gingras, Science 294, 1495 (2001).

[11] A. P. Ramirez, A. Hayashi, R.J. Cava, R. Siddharthan and B. S. Shastry, Nature 399, 333 (1999).

[12] X. Ke, B.G. Ueland, D. V. West, M. L. Dahlberg, R. J. Cava, and P. Schiffer, Phys. Rev. B 76, 214413 (2007).

[13] S. T. Bramwell, M. J. Harris, B. C. den Hertog, M. J.P. Gingras, J. S. Gardner, D. F. McMorrow, A. R. Wildes, A. L. Cornelius, J. D. M. Champion, R. G. Melko, and T. Fennell, Phys. Rev. Lett. 87, 047205 (2001).

[14] A. L. Cornelius and J. S. Gardner, Phys. Rev. B 64, 060406 (2001).

[15] B.C. den Hertog and M.J.P. Gingras, Phys. Rev. Lett. 84, 3430 (2000).

[16] D. Pomaranski, L.R. Yaraskavitch, S. Meng, K.A. Ross, H.M.L. Noad, H.A. Dabkowska, B.D. Gaulin, and J. B. Kycia, Nat. Phys. 9, 353 (2013).

[17] R.A. Borzi, F.A. Gómez Albarracín, H.D. Rosales, G.L. Rossini, A. Steppke, D. Prabhakaran, A.P. Mackenzie, D.C. Cabra, and S.A. Grigera, Nat. Commun. 7, 12592 (2016).

[18] X. Ke, M.L. Dahlberg, E. Morosan, J.A. Fleitman, R.J. Cava, P. Schiffer, Phys. Rev. B 78, 104411 (2008).

[19] H.D. Zhou, S.T. Bramwell , J.G. Cheng, C.R. Wiebe, G. Li, L. Balicas , J.A. Bloxsom, H. J. Silverstein, J.S. Zhou, J.B. Goodenough, and J.S. Gardner, Nature Comms. 2, 478 (2011).

[20] H.D. Zhou, J.G. Cheng, A.M. Hallas, C.R. Wiebe, G. Li, L. Balicas, J.S. Zhou, J.B. Goodenough, J.S. Gardner, and E.S. Choi, Phys. Rev. Lett. 108, 207206 (2012).

[21] J. G. A. Ramon, C. W. Wang, L. Ishida, P. L. Bernardo, M. M. Leite, F. M. Vichi, J. S. Gardner, and R. S. Freitas, Phys. Rev. B 99, 214442 (2019).

[22] R.D. Shannon, Acta Acta Crystallogr. A 32, 751 (1976).

[23] M.A. Subramanian, G. Aravamudan, and G.V. Subba Rao, Prog. Solid St. Chem. 15, 55 (1983).

[24] S. Nandi, Y.M. Jana, D. Swarnakar, J. Alam, P. Bag, and R. Nath, J. Alloys Compd. 714, 318 (2017).

[25] H. Liu, Y. Zou, L. Ling, L. Zhang, W. Tong, C. Zhangn, and Y. Zha, J. Magn. Magn. Mater. 369, 107 (2014).

[26] C. Castelnovo, R. Moessner and S. L. Sondhi, Nature 451, 42 (2008).

[27] C. P. Grams, M. Valldor, M. Garst, and J. Hemberger, Nature 5, 4853 (2014).

[28] S. R. Giblin, S. T. Bramwell, P. C. W. Holdsworth, D. Prabhakaran, and I. Terry, Nat. Phys. 7, 252 (2011).

[29] S. T. Bramwell1, S. R. Giblin, S. Calder, R. Aldus, D. Prabhakaran, and T. Fennell, Nature 461, 956 (2009).

[30] S. Scharffe, G. Kolland, M. Valldor 1, V. Cho, J.F. Welter, and T. Lorenz, J. Magn. Magn. Mater. 383, 83 (2015).

[31] P. Strobel, S. Zouari, R. Ballou, A.C. Rouhou, J.C. Jumas, and J.O. Fourcade, Solid State Sci. 12, 570 (2010).

[32] S. Nandi, Y.M. Jana, S. Sarkar, R. Jana, G.D. Mukherjee, and H.C. Gupta, J. Alloys Compd. 771, 89 (2019).

[33] S.T. Bramwell, M.N. Field, M.J. Harris, and I.P. Parkin, J. Phys.: Condens. Matter 12, 483 (2000).

[34] Sheetal, A. Ali, S. Rajput, Y. Singh, T. Maitra, and C.S. Yadav, J. Phys.: Condens. Matter 32, 365804 (2020).

[35] L.N. Mulay and E.A. Boudreaux, Theory and Applications of Molecular Diamagnetism, John Wiley and Sons (1976).

[36] A comparative statement of CW fit to dc susceptibility in different temperature zones for some Dy-based pyrochlores was made in Ref. [24]

[37] R. Siddharthan, B.S. Shastry, A.P. Ramirez, A. Hayashi, R.J. Cava, S. Rosenkranz, Phys. Rev. Lett. 83, 1854 (1999).

[38] V.K. Anand, D.A. Tennant, B. Lake, J. Phys.: Condens. Matter 27, 436001 (2015).

[39] S.K. Banerjee, Phys. Lett. 12, 16 (1964).

[40] B.Z. Malkin, T.T.A. Lummen, P.H.M. van Loosdrecht, G. Dhalenne, and A.R. Zakirov, J. Phys.: Condens. Matter 22, 276003 (2010).

[41] T. Yavors'kii, T. Fennell, M. J. P. Gingras, and S.T. Bramwell, Phys. Rev. Lett. 101, 037204 (2008).

[42] Q. Cui, Y.-Qi Cai, X. Li, Z.-Ling Dun, P.-Jie Sun, J.-Shi Zhou, H.-Dong Zhou, and J.-Guang Cheng, Chinese Phys. B 29, 047502 (2020).

[43] J. Snyder, J.S. Slusky, R.J. Cava, and P. Schiffer, Nature 413, 48 (2001).

[44] K. Matsuhira, Y. Hinatsu, and T. Sakakibara, J. Phys.: Condens. Matter 13, L737 (2001).

[45] J. Snyder, J.S. Slusky, R.J. Cava, and P. Schiffer, Phys. Rev. B 66, 064432 (2002); J. Snyder, B.G. Ueland, J.S. Slusky, H. Karunadasa, R.J. Cava, Ari Mizel, and P. Schiffer, Phys. Rev. Lett. 91, 107201 (2003).

[46] K. Matsuhira, Y. Hinatsu, K. Tenya, H. Amitsuka, and T. Sakakibara, J. Phys. Soc. Jpn. 71, 1576 (2002).

[47] J. Snyder, B. G. Ueland, A. Mizel, J. S. Slusky, H. Karunadasa, R. J. Cava, and P. Schiffer, Phys. Rev. B 70, 184431 (2004).

[48] J. Snyder, B. G. Ueland, J. S. Slusky, H. Karunadasa, R. J. Cava, and P. Schiffer, Phys. Rev. B 69, 064414 (2004).

[49] X. Ke, R. S. Freitas, B. G. Ueland, G. C. Lau, M. L. Dahlberg, R. J. Cava, R. Moessner, and P. Schiffer, Phys. Rev. Lett. 99, 137203 (2007).

[50] H.E. Stanley, Introduction to Phase Transitions and Critical Phenomena, Clarendon Press, Oxford, 1971.

[51] Y.M. Jana, A. Sengupta, and D.Ghosh, J. Magn. Magn. Mater. 248, 7 (2002).

[52] P. Henelius, T. Lin, M. Enjalran, Z. Hao, J. G. Rau, J. Altosaar, F. Flicker, T. Yavors'kii, and M. J. P. Gingras, Phys. Rev. B 93, 024402 (2016).

[53] G. C. Lau, R.S. Freitas, B.G. Ueland, B.D. Muegge, E.L. Duncan, P. Schiffer, and R. J. Cava, Nat. Phys. 2, 249 (2006).

[54] K. Baroudi, B. D. Gaulin, S. H. Lapidus, J. Gaudet, and R. J. Cava, Phys. Rev. B 92, 024110 (2015).

[55] B. G. Ueland, G. C. Lau, R. S. Freitas, J. Snyder, M. L. Dahlberg, B. D. Muegge, E. L. Duncan, R. J. Cava, and P. Schiffer, Phys. Rev. B 77, 144412 (2008).

[56] G. Sala, M. J. Gutmann, D. Prabhakaran, D. Pomaranski, C. Mitchelitis, J. B. Kycia, D. G. Porter, C. Castelnovo and J. P. Goo, Nature Mater. 13, 488 (2014).

[57] J.S. Gardner, A.L. Cornelius, L. J. Chang, M. Prager, Th. Brückel, G. Ehlers, J. Phys.: Condens. Matter 17, 7089 (2005).

[58] L. D. C. Jaubert and P. C.W. Holdsworth, J. Phys.: Condens. Matter 23, 164222 (2011).

[59] S. Hüfner, Optical Spectra of Transparent Rare Earth Components, Academic Press, New York, 1978.

[60] C.A. Morrison, R.P. Leavitt, Handbook on the Physics and Chemistry of Rare Earths, North-Holland, Amsterdam, 1982, Vol. 5, Chap. 46.

[61] M. Ruminy, E. Pomjakushina, K. Iida, K. Kamazawa, D.T. Adroja, U. Stuhr, and T. Fennell, Phys. Rev. B 94, 024430 (2016).

[62] G. Grimvall, Thermophysical Properties of Materials, North-Holland, Amsterdam, 1986.

[63] B. Klemke, M. Meissner, P. Strehlow, K. Kiefer, S.A. Grigera, and D.A. Tennant, J. Low Temp. Phys. 163, 345 (2011).

[64] S. Nandi, Y.M. Jana, and P. Bag, J. Chem. Thermodyn. 120, 174 (2018).